\documentclass[aps,prl,twocolumn,superscriptaddress,showpacs]{revtex4}

\usepackage{graphicx}
\usepackage{dcolumn}
\usepackage{bm}

\begin{document}
\preprint{APS/123-QED}

\title{Anomalous State Sandwiched between Fermi Liquid and Charge Ordered Mott-Insulating Phases of Ti$_{4}$O$_{7}$}

\author{M. Taguchi}

\affiliation{Soft X-ray Spectroscopy Lab, RIKEN SPring-8 Center, Sayo, Sayo, Hyogo 679-5148, Japan}

\author{A. Chainani}

\affiliation{Soft X-ray Spectroscopy Lab, RIKEN SPring-8 Center, Sayo, Sayo, Hyogo 679-5148, Japan}
\affiliation{Coherent X-ray Optics Lab, RIKEN SPring-8 Center, Sayo, Sayo, Hyogo 679-5148, Japan}

\author{M. Matsunami}

\affiliation{Soft X-ray Spectroscopy Lab, RIKEN SPring-8 Center, Sayo, Sayo, Hyogo 679-5148, Japan}

\author{R. Eguchi}

\affiliation{Soft X-ray Spectroscopy Lab, RIKEN SPring-8 Center, Sayo, Sayo, Hyogo 679-5148, Japan}

\author{Y. Takata}

\affiliation{Soft X-ray Spectroscopy Lab, RIKEN SPring-8 Center, Sayo, Sayo, Hyogo 679-5148, Japan}
\affiliation{Coherent X-ray Optics Lab, RIKEN SPring-8 Center, Sayo, Sayo, Hyogo 679-5148, Japan}

\author{M. Yabashi}

\affiliation{Coherent X-ray Optics Lab, RIKEN SPring-8 Center, Sayo, Sayo, Hyogo 679-5148, Japan}
\affiliation{JASRI/SPring-8, Sayo, Sayo, Hyogo 679-5198, Japan}

\author{K. Tamasaku}

\affiliation{Coherent X-ray Optics Lab, RIKEN SPring-8 Center, Sayo, Sayo, Hyogo 679-5148, Japan}

\author{Y. Nishino}

\affiliation{Coherent X-ray Optics Lab, RIKEN SPring-8 Center, Sayo, Sayo, Hyogo 679-5148, Japan}

\author{T. Ishikawa}

\affiliation{Coherent X-ray Optics Lab, RIKEN SPring-8 Center, Sayo, Sayo, Hyogo 679-5148, Japan}
\affiliation{JASRI/SPring-8, Sayo, Sayo, Hyogo 679-5198, Japan}

\author{S. Tsuda}

\affiliation{WPI-MANA, National Institute for Materials Science, Tsukuba 305-0044, Japan}

\author{S. Watanabe}

\affiliation{Institute for Solid State Physics, University of Tokyo, Kashiwa, Chiba 277-8581, Japan}

\author{C.-T. Chen}

\affiliation{Beijing Center for Crystal R$\&$D, Chinese Academy of Science, Zhongguancun, Beijing 100080, China}

\author{Y. Senba}

\affiliation{JASRI/SPring-8, Sayo, Sayo, Hyogo 679-5198, Japan}

\author{H. Ohashi}

\affiliation{JASRI/SPring-8, Sayo, Sayo, Hyogo 679-5198, Japan}

\author{K. Fujiwara}

\affiliation{Department of Advanced Materials Science, University of Tokyo, Kashiwa 277-8581 Japan }

\author{Y. Nakamura}

\affiliation{Department of Advanced Materials Science, University of Tokyo, Kashiwa 277-8581 Japan }

\author{H. Takagi}

\affiliation{Department of Advanced Materials Science, University of Tokyo, Kashiwa 277-8581 Japan }

\author{S. Shin}

\affiliation{Soft X-ray Spectroscopy Lab, RIKEN SPring-8 Center, Sayo, Sayo, Hyogo 679-5148, Japan}
\affiliation{Institute for Solid State Physics, University of Tokyo, Kashiwa, Chiba 277-8581, Japan}

\date{\today} 

\begin{abstract}
The Magn\'eli phase Ti$_{4}$O$_{7}$ exhibits two sharp jumps in resistivity with coupled structural transitions as a function of temperature at $T_{c1}$$\sim$142 K and $T_{c2}$$=$154 K. We have studied electronic structure changes across the two transitions using 7 eV laser, soft x-ray and hard x-ray (HX) photoemission spectroscopy (PES). Ti $2p$-$3d$ resonant PES and HX-PES show a clear metallic Fermi-edge and mixed valency above $T_{c2}$. The low temperature phase below $T_{c1}$ shows a clear insulating gap of $\sim$100 meV. The intermediate phase between $T_{c1}$ and $T_{c2}$ indicates a pseudogap coexisting with remnant coherent states. HX-PES and complementary calculations have confirmed the coherent screening in the strongly correlated intermediate phase. The results suggest existence of a highly anomalous state sandwiched between the mixed-valent Fermi liquid and charge ordered Mott-insulating phase in Ti$_4$O$_7$.
\end{abstract}


\maketitle

Ti$_4$O$_7$  is a member of the homologous series Ti$_n$O$_{2n-1}$ known as the Magn\'eli phase, and has been extensively studied over the past several decades because of its rich and puzzling properties\cite{mar72,lak76,ued02,eye04,leo06}. It is a mixed valence compound and exhibits strong anisotropy in electronic conduction\cite{ing83,ach03}, with chain structures observed in STM results\cite{nor98}. The system exhibits two first-order phase transitions in the temperature ($T$) dependence of the electrical resistivity $\rho(T)$ at $T_{c1}$$\sim$$142$ K and $T_{c2}$$=$154 K, while only one transition is observed in the magnetic susceptibility $\chi(T)$ at $T_{c2}$\cite{lak76}. 
In the low temperature (LT) phase below $T_{c1}$, it is  well-established that charge ordered chains of Ti$^{3+}$ are separated from each other by Ti$^{4+}$ chains. The $\rho(T)$ has a negative temperature coefficient and $\chi(T)$ almost vanishes in this phase, where the Ti $3d$ electrons are believed to be localized in Ti$^{3+}$-Ti$^{3+}$ pairs, stabilized by a bipolaron formation\cite{lak76,sch85}. In the high temperature (HT) phase above $T_{c2}$, there is no long-range order among the two types of Ti and the Ti valence is believed to be uniform 3.5$+$. The thermoelectric power and $\rho(T)$ are metallic with Pauli-paramagnetic $\chi(T)$\cite{sch85}. Thus, the HT transition at $T_{c2}$ is attributed to a delocalization of the $3d$ electrons.
Nonetheless, recent photoemission spectroscopy (PES) studies of Ti$_{4}$O$_{7}$ have essentially revealed only the absence of a Fermi-edge in the HT phase\cite{abb95,kob02}. The extremely broad feature and no spectral weight at $E_F$ in the metallic state reflect a completely incoherent motion, which is in contradiction to early transport measurements indicating a Fermi liquid state. This is a key controversial issue in this material. 

Another contradiction is related to the nature of electronic states in the intervening region between $T_{c1}$ and $T_{c2}$, which has been a subject of debate for several decades  and its understanding is far from satisfactory.  
With decreasing $T$ from HT phase, the $\rho(T)$ increases by three orders of magnitude at $T_{c2}$ and shows a negative temperature coefficient. The $\chi(T)$ almost vanishes at $T_{c2}$ as in the LT phase. Previous x-ray diffraction studies provided  no direct evidence for a transition from charge localization to metal formation at $T_{c2}$, but anomalously large thermal factors were observed\cite{mar73}. 
One of the most commonly accepted models for describing the LT phase is the Verwey model\cite{ver41}  with charge ordering of Ti$^{3+}$ and Ti$^{4+}$. 
The intermediate temperature (IT) phase was viewed as a partially ordered bipolaron liquid state\cite{lak76,sch85}. Very recently, a twist to this debate was provided by PES experiments\cite{kob02} asserting that IT phase may be described by a soft-Coulomb gap (SCG) model\cite{efr75}, with localized Ti $3d$ states near Fermi level ($E_F$).
However, the structure study shows appearance of a fivefold superstructure in the IT phase\cite{pag84}, which can imply the presence of long-range ordering. Clearly this result goes beyond the SCG and bipolaron liquid scenario  with dynamical disorder. Therefore, no conclusive picture has emerged yet for the nature of IT phase.

\begin{figure}
\includegraphics[scale=.62]{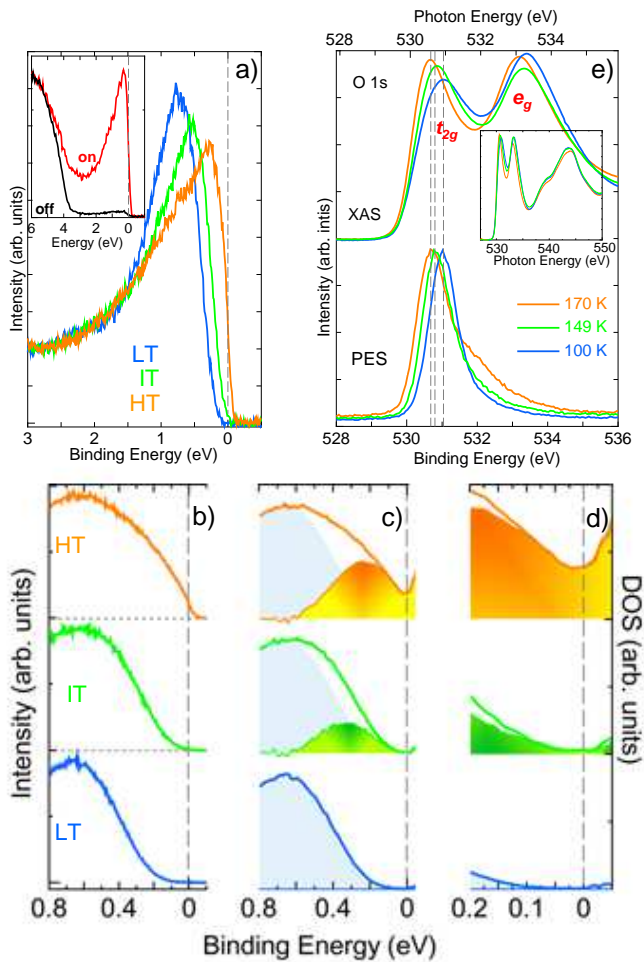}
\caption{\label{fig1}   
(Color online)  $T$-dependent electronic structure near $E_F$. (a) Resonant PES at the Ti $L_3$ edge. The inset shows the on- and off- spectra at HT-phase over a wide energy range. (b) Experimental Laser-PES spectra. (c) The spectral function $A(\omega)$. Shaded areas (yellow and green) highlight the difference of $A(\omega)$ to the LT-phase $A(\omega)$ (light blue). (d) Enlarged plot of $A(\omega)$ near $E_F$. (e) The O 1s core-level shift of HX-PES compared to XAS. Upper panel, O $1s$ XAS. Lower panel, O $1s$ HX-PES. Wide-range O $1s$ XAS are shown in the inset. }
\end{figure}

In this Letter, we critically reexamined the nature of electronic states near $E_F$ and addressed two fundamental issues mentioned above. We employed various spectroscopies using a wide range of incident photon energies ($h\nu$$=$$7$ eV $\sim 8$ keV) such as core-level, off- and on- resonant valence band PES and x-ray absorption spectroscopy (XAS). We show that the HT state can be well described as a mixed-valent Fermi Liquid. We also found an anomalous electronic state: no gap, no Fermi-edge, but with remnant coherent states near $E_F$ in the IT phase. This behavior is at odds with the existing models. 

UV-laser excited (Laser)- PES measurements were performed using a Scienta R4000 electron analyzer and an ultra-violet ($h\nu$$=$$6.994$ eV) laser for the incident light\cite{kis08}. The base pressure of the chamber was below $\sim$$5$$\times$$10^{-11}$ Torr throughout the measurements. The energy resolution was $\Delta E\sim5.0$ meV. 
Hard x-ray (HX)-PES was performed using a photon energy $h\nu$$=$$7.93$ keV, at a vacuum of $1$$\times$$10^{-10}$ Torr. The measurements were carried out at undulator beam line BL29XUL, SPring-8 using a Scienta R4000-10KV electron analyzer\cite{tam01}. Soft x-ray (SX)-PES was performed at BL17SU. The total energy resolution, $\Delta E$ was set to $\sim 0.2$ eV for both SX- and HX-PES measurements. The O $1s$ XAS was obtained in total electron yield mode. Single crystals of Ti$_4$O$_7$ were grown by the floating zone technique. In all measurements, clean sample surfaces were prepared by fracturing $in$-$situ$ and gold $E_F$ was measured to calibrate the energy scale.

First, we present the on- and off- resonant PES spectra of the HT phase for a wide binding energy ($E_B$) region in the inset of Fig.~1(a). The off-resonance spectrum with $h\nu$$=$$450$ eV shows a dominant contribution of the O $2p$ states to valence band peak at binding energies of $\sim$6 eV, while weak Ti $3d$ feature was observed near $E_F$ as in earlier studies\cite{abb95,kob02}. On-resonance data in HT phase has shown a strong enhancement of the Ti $3d$ feature and a clear Fermi-edge. This is because the narrow Ti $3d$ band just below $E_F$ is selectively magnified by tuning x-ray photon energy to the maximum of Ti $L_3$ XAS white line ($h\nu$$=$$460$ eV). 
The $T$-dependent resonant-PES spectra near the $E_F$ are shown in Fig.~1(a). 
On decreasing $T$, the spectral weight at $E_F$ has transferred to the higher binding energy region, forming a gap ($\sim$100 meV) in the LT insulator phase. 
The spectral weight transfer occurs from $E_F$ up to an energy scale of 1 eV and is indicative of a correlation driven Mott insulating phase\cite{ima98}.
In the IT phase, however, the spectrum has shown neither a Fermi-edge nor a real gap, and is the central issue in this letter as discussed in the following. 

To make sure whether there remains a well-defined residual spectral weight at $E_F$ in the IT phase, we carried out Laser-PES measurements with  high energy resolution. 
The key observation here is that the IT phase spectrum showed neither a Fermi-edge nor a real gap even with a high resolution of  5 meV.
The $T$-dependent Laser-PES data are shown in Fig.~1(b). While the present results are seemingly similar to the results using He lamp presented previously\cite{kob02}, the greatly improved energy resolution in the current study allows us for the first time to finely resolve the electronic structure near $E_F$. 
Nevertheless, we could not observe any evidence of gap opening in IT phase. One may possibly attribute this gapless feature to the occupancy by thermally excited electrons, because of the rather high temperature ($\sim$140 K). In order to extract the intrinsic $T$-dependence of one electron spectral function $A(\omega)$, we eliminate the effect of the Fermi-Dirac function $f(\omega)$. We evaluate $A(\omega)$ by using the general expression $I(\omega)=\int d\epsilon A(\omega-\epsilon)f(\omega-\epsilon)g(\epsilon)$, where $g(\epsilon)$ is the Gaussian broadening corresponding to the instrumental resolution (5.0 meV). As a results, we obtained $A(\omega)$ not only below but also above $E_F$ (within $E_{B}$ $\sim$ 5k$_{B}T$), owing to the occupancy by thermal excitations. The $T$-dependent evolution of the extracted $A(\omega)$ on a wider energy range ($-$0.05 eV $\le$ $E_B$ $\le$ 0.8 eV) and in the vicinity of the $E_F$ are shown in Fig.~1(c) and 1(d), respectively. The yellow and green shaded areas are the difference spectra obtained by subtracting the LT phase $A(\omega)$ from HT and IT $A(\omega)$, respectively. Three distinct spectral shapes were observed: (i) The LT phase has a gap of $\sim$100 meV followed by a broad peak at ~0.7 eV. (ii) The presence of the Fermi-edge in HT phase is still clear although it is smaller in magnitude compared to resonance data due to the cross section. (iii) The IT phase $A(\omega)$ exhibited a definite but small residual spectral weight at $E_F$. This is the most important observation emphasized here. The pseudogap observed in the present PES spectra is similar to the suppression of the low-frequency spectral weight in optical studies\cite{wat07}. This clearly shows the existence of an electronic state between $E_F$ and 0.6 eV binding energy in IT phase. As we will discuss later, we attribute this new state to a coherent screening state. 

\begin{figure}
\includegraphics[scale=.60]{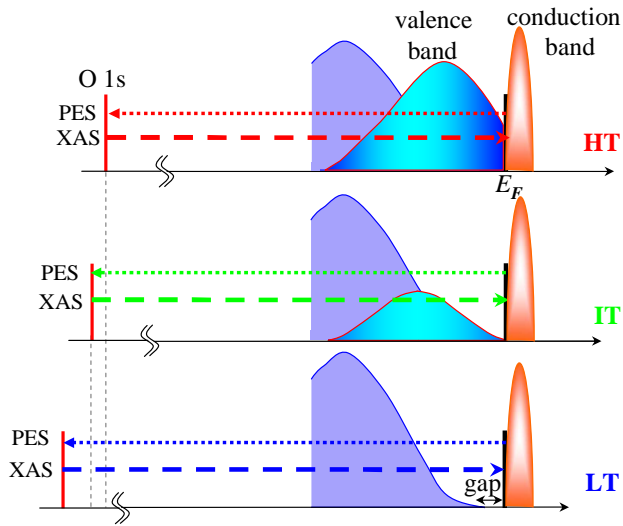}
\caption{\label{fig2}     
(Color online) Schematic pictures illustrating the evolution of the low energy electronic structure with temperature. }
\end{figure}

We so far discussed on the occupied states below $E_F$. It is, however, important to consider that the real gap is also affected by the unoccupied state above $E_F$. To this end, we carried out O $1s$ XAS measurement, shown in upper panel of Fig.~1(e).  XAS is a complementary probe of the unoccupied state of a material, providing site and symmetry projected density of states. The O $1s$ XAS spectra exhibit a double-peaked sharp structure near the pre-edge region between 529 and 536 eV and a broader structure around 536-550 eV\cite{abb95}. The data presented here are very similar to that reported earlier\cite{abb95}. The double-peaked structure has been identified to O $2p$ states, which are strongly hybridized with unoccupied Ti $3d$ $t_{2g}$- and $e_g$- orbital. 

A systematic shift of the $t_{2g}$ pre-edge peak at $\sim$531 eV was clearly observed as a function of $T$. This $t_{2g}$ peak in IT (LT) phase has exhibited a shift of 100 meV (300 meV) in peak position towards higher energy with respect to HT phase spectrum, while the 100 meV shift in IT phase was missing in earlier result\cite{abb95}. We would like to emphasize the importance of this observation. These shifts directly indicate that the lowest energy state of the unoccupied state for IT and LT phase are 100 and 300 meV higher in energy than that of HT phase from the O $1s$ level, respectively (see Fig.~2). 
On the other hand, same amount of the peak shifts were also observed in O $1s$ core-level HX-PES (see Fig.~1(e)) which reflect the $E_F$ shifts with respect to the O $1s$ level, as shown in Fig.~2. Therefore, we can definitively conclude that the $E_F$ always resides near the bottom of the conduction band, resulting in no gap above $E_F$ for all three phases.
 This conclusion is consistent with the previous study on the Seebeck coefficient $S$ indicating that the mobile carrier is $n$-type\cite{sch85}.

Further information on the nature of the pseudogap state in IT phase can be obtained from the bulk sensitive HX-PES. This technique has enabled us to obtain clear evidence of the coherent screening (well-screening) due to electronic states at and near $E_F$. 
The combination with extended configuration interaction model (CIM) calculation can provide the clear evidence of the modulation of the bulk electronic state near $E_F$\cite{hor04,tagprb05,tagprl05,tagprl08}. 
Figure~3 shows a complementary set of experimental and calculated Ti $2p$ core level spectra for the three phases. 
The experimental spectra in Fig.~3(a) were normalized for area under the curve. 
An extremely unusual $T$-dependence was observed with clear changes across $T_{c1}$ and $T_{c2}$. In the following we analyze the spectral changes in detail.

\begin{figure}
\includegraphics[scale=.47]{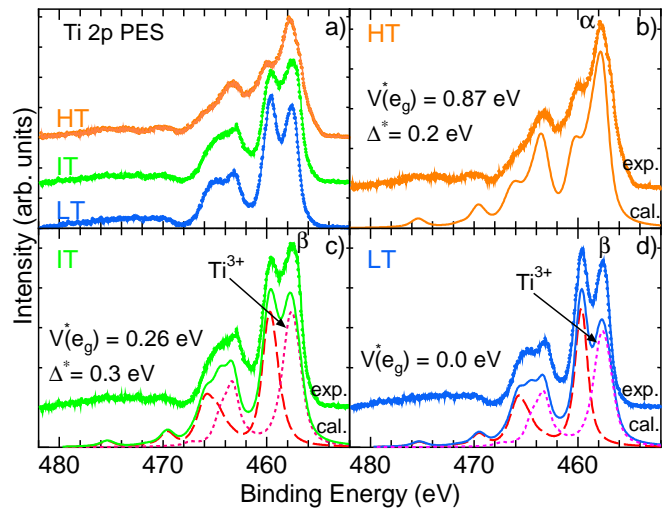}
\caption{\label{fig3}     
(Color online)  (a) Comparison between  experimental Ti $2p$ HX-PES spectra for HT-, IT- and LT-phases. (b), (c), (d) CIM calculations (lower panel) are compared with experiments. The doted curves are the Ti$^{3+}$ component. }
\end{figure}

The Ti $2p$ HX-PES spectra were calculated within the extended CIM with $C_{3v}$ local symmetry. This model is well established and was successfully used to study the PES spectra for various materials. Details of the model have been described in previous work\cite{hor04,tagprb05,tagprl05,tagprl08}. We used, as basis states, six configurations: $3d^0$, $3d^{1}\underline{L}$, $3d^2\underline{L}^2$, $3d^1\underline{C}$, $3d^2\underline{C}^2$ and $3d^2\underline{CL}$. The $3d^1\underline{C}$ represents the charge transfer (CT) between Ti $3d$ and the coherent state at $E_F$, labeled $C$. An effective coupling parameter $V^*$, for describing the interaction strength between the Ti $3d$ and coherent state is introduced, analogous to the Ti $3d$$-$O $2p$ hybridization $V$. The parameter values used are (in eV): on-site  Coulomb repulsion $U$$=$$4.5$, the attractive core-hole potential $U_{c}$$=$$5.5$, the CT energy $\Delta$$=$$5.0$, the crystal field $10Dq$$=$$0.7$, the trigonal crystal field $\Delta_{trg}$$=$$-0.05$, $V(e_g)$$=$$2.9$. Theory has reproduced the experiments very satisfactorily for all phases, as shown in Fig.~3 (b)-(d). 

For a finer comparison between theory and experiment let us first consider the sharp peak labeled $\alpha$ of HT phase in Fig.~3(b), where the final state is well screened $\underline{2p}3d^1\underline{C}$ state. The screening effect from the coherent state $C$ near $E_{F}$ leads to the formation of the low energy peak $\alpha$. In addition, the ground state of HT phase mainly consists of 35.4\% $3d^0$, 32.6\% $3d^{1}\underline{L}$, and 15.4\% $3d^1\underline{C}$, indicating strong mixed-valent ground state.
Next, we consider the LT phase spectrum. Because of the complete absence of the coherent state near $E_F$ in LT insulating phase (see Fig.~1(c)), we set $V^*(e_g)$$=$$0$, leading to a complete suppression of the $\alpha$ feature. The corresponding spectrum is shown by dashed line in Fig.~3(d).
The calculated spectrum does not account for the lowest binding energy feature, labeled $\beta$, which occurred 0.2 eV below the $\alpha$ feature. The difference matches rather well with the calculated spectrum for Ti$^{3+}$, shown by dotted line in Fig.~3(d).
 This confirms the existence of Ti$^{3+}$ in LT phase, as is well established from various x-ray diffraction studies. The total calculated spectrum is obtained by a linear combination of CIM and Ti$^{3+}$ states with a relative weight of 50 \% and 50 \%, respectively. The agreement is remarkable (see Fig.~3(d)). Our calculations basically confirmed the standard valence assignment: the mean valence for Ti is approximately $3.5$ in HT phase, and clear spectral signatures of Ti$^{3+}$ and Ti$^{4+}$ in the LT phase.

Finally, we discuss the IT phase spectrum. Since XRD studies suggest the existence of Ti$^{3+}$ in IT phase as well as LT phase, the peak $\beta$ is predominantly due to the Ti$^{3+}$ derived state. However, there is a definite enhancement of the peak $\beta$ compared to the LT spectrum. We attribute this enhancement to the screening from the small amount of the coherent state, as shown in the shaded areas in the middle panel of Fig.~1(c). In fact, good agreement is obtained between the experiment and the Ti$^{3+}$ $+$ CIM calculation with the small $V^*(e_g)$ spectra. 
We therefore conclude that the electronic state within the gap in IT phase has a coherent character. 
The present results together with the previous superstructure observation suggest that the interpretation beyond the bipolaron liquid and SCG picture is necessary in the IT phase.

In conclusion, we have investigated the evolution of the electronic structure of Ti$_{4}$O$_{7}$ with $T$ by using the various spectroscopy and addressed two important issues. Figure~2 summarizes our main results. From O 1s XAS and HX-PES, the any gap formation above $E_F$ was not observed. Thanks to the sensitivity of resonant PES to the Ti $3d$ state, the Fermi-edge was observed clearly in HT metal phase, suggesting the Fermi Liquid phase. In the intervening region between the HT Fermi Liquid and LT charge ordered phases, the pseudogap feature (no-gap, no-Fermi-edge) was observed in Laser-PES.  Contrary to the SCG model, Ti $2p$ HX-PES results have indicated that this pseudogap feature is associated with the coherent screening states.  

M. T. would like to thank A. Fujimori for useful information and discussions. 
This experiment with soft x-ray was carried out with the approval of the RIKEN SPring-8 Center (Proposal No. 2008197). This work was partially supported by KAKENHI (20540324).

\end{document}